\documentclass[12pt,preprint]{aastex}

\shorttitle{Astrophysical production of black holes}
\shortauthors{Barrau et al.}

\begin{document}

\title{Astrophysical production of microscopic black holes\\
    in a low Planck-scale World}

\author{A. Barrau, C. F\'eron \& J. Grain}
\affil{Laboratory for Subatomic Physics and Cosmology, CNRS-IN2P3\\ 
Joseph Fourier University, Grenoble, France}

\email{Aurelien.Barrau@cern.ch}

\begin{abstract}
In the framework of brane-world models lowering the Planck scale to the TeV 
range, it has recently been pointed out that small black holes could be formed 
at particle colliders or by neutrinos interactions in 
the atmosphere. This article aims at reviewing other places and epochs 
where microscopic black holes could be formed~: the interstellar medium and the
early Universe. The related decay channels and
the propagation of the emitted particles are studied to conclude that, in spite
of the large creation rate for such black holes, the amount of produced 
particles do not conflict with experimental data. This shows, from the
astronomical viewpoint, that models with large extra dimensions making the gravity
scale much lower are compatible with observations.

\end{abstract}


\section*{Introduction}
It has recently  been  shown  that black holes could be  formed at
future colliders if  the  Planck  scale  is of order a TeV, as is the case in
some extra-dimension  scenarios \citep{dimo,gidd}. This idea has driven
a considerable amount of interest : see  {\it  e.g.} 
\citep{idea1,idea3,idea4,idea5,idea6,idea7,idea8,idea2}. The
same  phenomenon could also occur (and could be detected) due to
ultrahigh  energy  neutrino interactions in the atmosphere 
\citep{ancho3,ancho-cr,ancho1,ancho2,ancho4,ancho5}.
Such possibilities open very exciting ways to search for new physics, {\it
e.g.} quantum-gravitational effects through a Gauss-Bonnet term in the action
\citep{newphys1}, Higgs production \citep{newphys2}, evaporation into 
supersymmetric particles \citep{newphys3} or a D-dimensional cosmological 
constant \citep{newphys4}. This article aims at clarifying the
situation about the possible consequences of an astrophysical and cosmological
production of microscopic black holes ($\mu$BH) in the Universe. This is
especially important as it has been argued that a low Planck scale could
already be excluded by the available data on galactic cosmic-rays.
The first section summarizes
the general framework of TeV Planck scale models and the basics of cross
section computations. Section 2 is devoted to the investigation of black hole
production by the interaction of high energy cosmic-rays with the
inter-stellar medium (ISM). The subsequent decay of those black holes
and the propagation of the emitted particles are also studied. 
Section 3 deals with a few other exotic possibilities, including black hole 
relics and thermal creation of black holes in the early Universe. Finally, 
we conclude that scenarios lowering the Planck scale in the TeV era are viable
from the astrophysical point of view.

\section{TeV Planck scale framework}

The "large extra  dimensions" scenario \citep{large1,large2,large3} is a very elegant
way to address geometrically the  hierarchy  problem  (among  others),
allowing only the gravity to propagate in the bulk. The Gauss law relates
the Planck  scale of the  effective 4D low-energy theory $M_{Pl}$ with
the  fundamental  Planck   scale  $M_D$  through  the  volume  of  the
compactified dimensions, $V_{D-4}$, via:
$$M_{D}=\left(\frac{M_{Pl}^2}{V_{D-4}}\right)^{\frac{1}{D-2}}.$$
It is thus possible  to set $M_D\sim$~TeV without being  in  contradiction
with any currently available
experimental data. This translates into radii values 
between a  fraction of a
millimeter and a  few  Fermi for  the  compactification radius of  the
extra dimensions (assumed to  be of same size and flat, {\it  i.e.} of
toroidal shape). Furthermore, such a small value for the Planck energy
is not artificial and
could be somehow expected so as to minimize the difference  between the electroweak
and Planck  scales, as motivated by the construction of this approach.
In such a scenario, at sub-weak energies, the standard model (SM) fields 
must be localized to a 4-dimensional manifold of weak scale "thickness" 
in the extra dimensions. As shown in \citep{large1,large2, large3}, as an example based 
on a dynamical assumption with D=6, it is possible to build such a 
SM field localization. This is however the non-trivial task of those
models.\\

Another  important  way for realizing TeV scale  gravity  arises  from
properties   of   warped   extra-dimensional   geometries   used    in
Randall-Sundrum scenarios \citep{rs}. If the  warp  factor  is small in
the vicinity of the standard model brane, particle masses can take TeV
values, thereby  giving rise to a large hierarchy  between the TeV and
conventional  Planck  scales  \citep{gidd,katz}.  Strong  gravitational
effects  are  therefore  also  expected  in   high  energy  scattering
processes on the brane.\\

In those  frameworks, black holes could be formed  by any interaction
above the Planck scale.  Two elementary particles with a center-of-mass energy  $\sqrt{s}$
moving in  opposite directions with  an impact parameter less than the
horizon  radius  $r_+$  should  form  a  black  hole  of mass  $M  
\approx \sqrt{s}$ with a cross  section expected  to  be of  order
$\sigma \approx  \pi r_+^2$. Those values are in fact approximations as
the black hole mass will be only a fraction of the center-of-mass energy 
whose exact value depends on the dimensionality of the space-time and 
on the angular momentum of the produced black hole. A significant part of the
incoming energy is lost to gravitational radiation. In particular, lower limits
on the inelasticity (ratio of the mass of the formed black hole to the incoming
parton energy) where derived in \citep{yosh1} as a function of the impact
parameter. Although those bounds where slightly increased in \citep{yosh2}, they
still can be considered as reasonable estimates of the consequences of the
inelasticity and where
shown in \citep{ancho-in} to drastically reduce the expected number of produced
black holes. To remain conservative (and except otherwise stated) the
computations performed in this article use a unity inelasticity to maximise the
emitted flux and strengthen the conclusions.\\

To compute  the real probability  to form
black holes in a nuclei collision, it is necessary to take into account that only
a fraction  of the total center-of-mass energy is  carried out by each
parton  and  to   convolve  the  previous  estimate  with  the  parton
luminosity  \citep{dimo}. The proton-proton cross section can be written as
$$
    \frac{d\sigma(pp \to \mbox{BH} + X)}{d M_{BH}}=
    \sigma(ij \to \mbox{BH})
    \left|_{\sqrt{s}=M_{BH}}\right.\frac{2 M_{BH}}{s} 
    \sum_{i,j} \int_{M^2_{BH}/s}^1  
    \frac{dx_i}{x_i} f_i(x_i) f_j(\frac{M^2_{BH}}{s x_i}),
$$

where $f_a(x_a)$ are the parton distribution functions (PDFs) and
$$
    \sigma(ij \to \mbox{BH}) \approx \pi r_+^2 = \frac{1}{M_{Pl}^2}
    \left[ 
    \frac{M_{BH}}{M_{Pl}} 
      \left( 
        \frac{8\Gamma\left(\frac{D-1}{2}\right)}{D-2}
      \right)
    \right]^\frac{2}{D-3}.
$$\\

Those PDFs can be inferred from \citep{pdf} or rescaled from \citep{dimo} at various
center-of-mass energies by taking advantage of the fact that they only depend on
$y=M_{BH}/\sqrt{s}$ after an appropriate change of variables.\\

Up to now, the different works related with those phenomena have focused on the
production of $\mu$BHs at colliders or through neutrino interactions in the
earth atmosphere. The first approach is extremely promising as it would 
allow those
fascinating objects to be formed and to evaporate within the largest and more
sophisticated detectors ever built by human beings. Many new physics effects
could be discovered, related with the dimensionality of space-time, with the
radii of extra-dimensions, with new particles, with quantum gravity, etc. The
second approach is based on the higher center-of-mass energy available
when extremely high energy neutrinos (allowing for a high signal to noise ratio
due do the very small standard model cross sections) interact with atmospheric 
nuclei.\\

On the other hand, there are other places in the Universe where one should
expect $\mu$BHs to be formed if the Planck scale lies in the TeV range. Namely, the
ISM where high energy cosmic-ray interactions occur above this threshold and the 
very early Universe where the temperature could have been arbitrarily high. The
specific energy and time scales associated with those particular conditions
deserve a dedicated study.

\section{Black hole production by cosmic-ray interactions on the ISM}

The number of black holes produced by cosmic-rays on the interstellar medium
per unit of time, per unit of volume and per
unit of mass can be written as
$$\frac{dN}{dM_{BH}dtdV}(M_{BH})=
\int_{\frac{M_D^2}{2M_{pl}}}^{\infty}{4\pi~\phi(E_{CR})~n~\frac{d\sigma}{dM_{BH}}(M_{BH},E_{CR})
dE_{CR}}$$
where $n$ is the ISM proton density and $\phi(E_{CR})$ is the cosmic-ray
differential spectrum at energy $E_{CR}$. Assuming that the ISM is made of 90\%
of hydrogen and 10\% of helium, $n$ can be taken at $\sim 1.3$~cm$^{-3}$. As we
deal with a diffusion process insensitive to local overdensities or underdensities,
this average approximation holds very well. For 
$M_D\sim 1$~TeV, which is the order of magnitude of the lowest Planck scale value
compatible with experimental data (see references in \citep{kanti}), the threshold
energy for cosmic-rays is $5.5\times 10^{14}$~GeV, which lies just below the knee. 
Measurements from the KASKADE experiment \citep{kaskade} where used to evaluate
$\Phi(E_{CR})$, taking into
account the
different contributions from hydrogen, helium, carbon, silicon and iron. The
cross sections have been derived, following \citep{dimo}, by the method given in
section 2.
Table
\ref{tab1} gives the resulting differential numbers of $\mu$BHs produced and their
initial temperatures as a function of their mass for two different Planck scales and
two extreme numbers of dimensions ($D=5$ is not included as a very low
Planck scale with only one extra dimension is already excluded on the ground of the
dynamics of the solar system).\\

It can be seen that the production rate is small, due to the low value of the flux in
this PeV-EeV range. It is also very sensitive to the $\mu$BH mass because the 
cross section for $p+p\rightarrow BH$ is a fast decreasing function of $M_{BH}$.  Nevertheless,
when the volume of our Galaxy $\sim10^{60}$~m$^3$ is taken in account this leads to
$\sim 10^{19}$ $\mu$BHs created per second within the Milky Way for $M_D=1$~TeV.
The D-dependence is
very weak and not included in the table as uncertainties clearly dominate at this
level of accuracy.\\

\begin{table}[!hhh] 
\begin{center}
\begin{tabular}{|c|c|c|c|c|}
\hline
$M_D$~\small (GeV) & $M_{BH}$~\small(GeV) &
$\frac{dN}{dM_{BH}dtdV}$~\footnotesize(${\rm GeV}^{-1}.s^{-1}.m^{-3}$) &
D & $T_{BH}$~\small(GeV)\\
\hline \hline
1 Tev & $M_D$ & $ 10^{-41}$ & 11 & 434 \\
~~~ & ~~~ & ~~~ & 6 & 172 \\
\hline
~~~ & 10$M_D$ & $ 10^{-45}$ & 11 & 325 \\
~~~ & ~~~ & ~~~ & 6 & 80 \\
\hline
~~~ & 100$M_D$ & $ 10^{-50}$ & 11 & 244 \\
~~~ & ~~~ & ~~~ & 6 & 37 \\
\hline \hline
10 Tev & $M_D$ & $ 10^{-45}$ & 11 & 4340 \\
~~~ & ~~~ & ~~~ & 6 & 1725 \\
\hline
~~~ & 10$M_D$ & $ 10^{-50}$ & 11 & 3258 \\
~~~ & ~~~ & ~~~ & 6 & 800 \\
\hline
~~~ & 100$M_D$ & $ 10^{-55}$ & 11 & 2443 \\
~~~ & ~~~ & ~~~ & 6 & 371 \\
\hline
\end{tabular}
\caption{Production rate and initial temperature of $\mu$BHs for different numbers of
dimensions, different Planck scales and different masses.}
\label{tab1}
\end{center}
\end{table}

\begin{figure}
\epsscale{.80}
\plotone{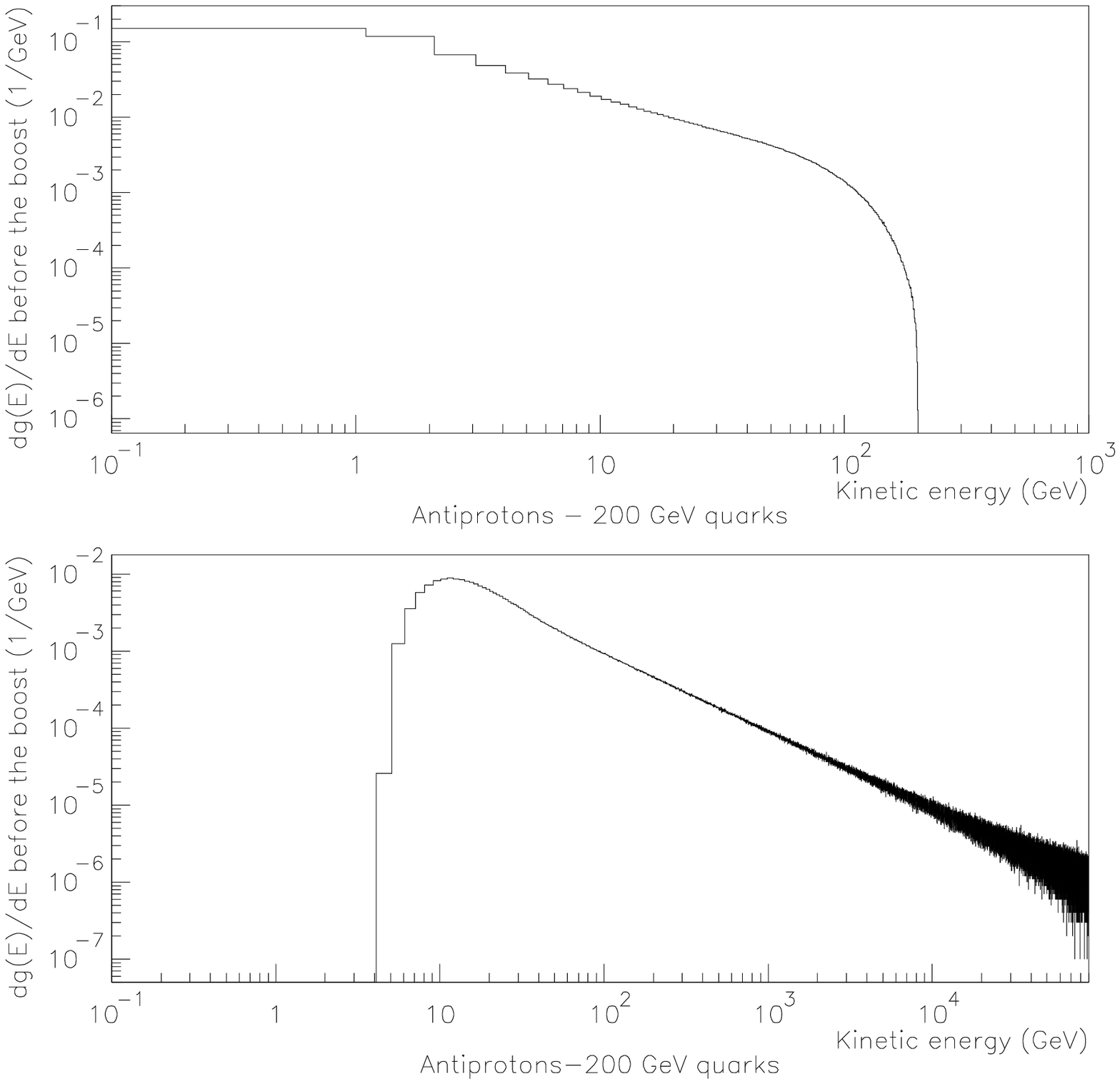}
\caption{Fragmentation function into antiprotons for jets of u quarks emitted by a
200 GeV black hole produced by a $10^8$~GeV cosmic-ray as a function of the 
antiproton kinetic energy. Upper part~: in the $\mu$BH frame. Lower part~: in the galactic frame.\label{fig1}}
\end{figure}

To investigate the experimental consequences of those possibly formed black holes,
their evaporation should be taken into account. As already studied for 
4-dimensional primordial black holes \citep{pbar,pbar2,pbar3} antiprotons are very promising
for this purpose as their abundance is both well known and very small when compared
to matter cosmic-rays ($\bar{p}/p\sim 10^{-5}$) \citep{fio}. The source term for
this process can be written as:

$$\frac{dN_{\bar{p}}}{dE_{\bar{p}}dtdV}=\int_{M_{BH}=M_D}^{\infty}
\int_{E_{CR}=\frac{M}{2m_p}}^{\infty}\sum_i
\frac{d\sigma}{dM_{BH}}(M_{BH},E_{CR})$$
$$\times~n\Phi(E_{CR})\varphi(M_{BH})  N_i \frac{dg_i}{dE_{\bar{p}}}(E_{\bar{p}},M,E_{CR})
dE_{CR}dM_{BH}$$

where $\varphi$ stands for the number of emitted quanta by a black hole of
mass $M_{BH}$ and temperature $T_{BH}$~: $\varphi \sim M_{BH}/(2T_{BH}) \sim 2\pi r_+M_{BH}/(D-3)$. The
relative number of quarks or gluons of type $i$ is $N_i$ and the fragmentation 
function for such a parton of energy $T_{BH}$ into an antiproton of kinetic
energy $E_{\bar{p}}$ is given by $dg_i/dE_{\bar{p}}$. This term also depends on $E_{CR}$ 
as the produced $\mu$BH is Lorentz boosted with respect to the galactic frame. 
It has been determined using the Lund model PYTHIA \citep{pythia}
Monte-Carlo generator, between the thresholds
$E_{min}\approx\gamma_{BH}T_{BH}(1-\beta_{BH})$ and 
$E_{max}\approx\gamma_{BH}T_{BH}(1+ \beta_{BH})$, as shown on the example of
Fig.~\ref{fig1} where the important r\^ole played by the center-of-mass velocity can
be easily seen.

\section{Decay and related spectra}

Once this source term is established, the produced antiprotons should be allowed to
propagate within the Galaxy. For this purpose, a two zone diffusion model described 
in \citep{fio} \& \citep{david} have been used.
In this approach, the geometry of the Milky-Way is a cylindrical box embedded in
a diffusion halo  whose extension is still subject to 
large uncertainties. The five parameters used are~: $K_0$,
$\delta$ (describing the diffusion coefficient $K(E)=K_0 \beta 
R^{\delta}$), the
halo half height L, the convective velocity $V_c$ and the Alfv\'en velocity
$V_a$. They have been varied within a given range determined by an exhaustive and
systematic  study of cosmic-ray nuclei data \citep{david} and chosen at their mean value. 
The spectrum is affected by energy
losses when antiprotons interact with the galactic interstellar matter
and by energy  gains when reacceleration occurs.
These energy changes are described by an intricate integro--differential equation
\citep{pbar2} where a source term $q_{i}^{ter}(E)$ was added, leading to the
so-called tertiary component which corresponds to inelastic but 
non-annihilating reactions of $\bar{p}$ on interstellar matter.\\

\begin{figure}
\epsscale{.80}
\plotone{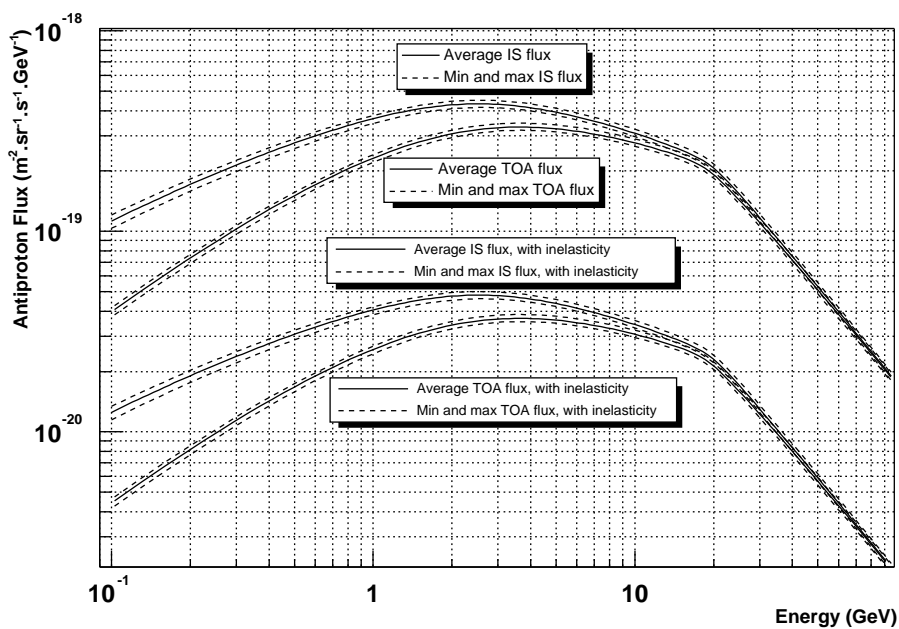}
\caption{ Antiproton spectra due to the evaporation of $\mu$BHs produced
by cosmic-ray interactions on the ISM. The upper set of curves is an upper limit
ignoring inelasticity and the lower set of curves takes into account the
inelasticity effect as estimated in \citep{ancho-in}. In each case, the
interstellar and "top of atmosphere" fluxes are plotted. The dotted lines are the
minimum and maximum spectra (when varying astrophysical parameters) and the
solid line is the mean spectrum.\label{fig2}}
\end{figure}

Figure \ref{fig2} shows the resulting antiproton spectrum around its maximum.
The different curves correspond to different astrophysical parameters and to
the interstellar and "top of atmosphere" fluxes. The inelasticity has also been taken
into account \citep{ancho-in} as a simple resclaling as those results are extremely
weakly dependant upon the details of the shape of the mass spectrum. In any
case, they remain extremely far below the background which lies at the level of a few
$10^{-2}$~GeV$^{-1}$sr$^{-1}$s$^{-1}$. But it should be pointed out that
the spectrum is much harder than the background. The reasons for this are quite
straightforward. First, although quite high for an exotic process, the cross section
for $\mu$BHs production is small when compared to SM processes : 
it is of the order of a few hundreds picobarns at the
threshold whereas the total $pp$ cross section 
is not far from 100~mb with an antiproton multiplicity close to one for
$\sqrt{s}\sim 1$~TeV. This explains the overall normalization around the
maximum. Furthermore, the antiproton flux results from the convolution of the 
primary cosmic-ray spectrum with  the fragmentation function. For the secondary 
component (antiprotons due du $pp$ interactions), this 
fragmentation function is not far from a Dirac distribution
and the spectrum is expected to have roughly the same slope than the primary one.
For the $\mu$BH
component considered in this article, the fragmentation function is
substantially widened by the boost and becomes the hardest function of the 
convolution equation. The resulting spectrum therefore exhibits the same slope and
becomes much harder than for the secondary component. The confinement effects 
due to the galactic magnetic field induce
a softening of the spectra below the knee but the spectral index difference 
remains the same whatever the energy. Due to this important slope difference,
the antiproton flux due to $\mu$BHs should dominate
over the secondary one around $4\times 10^{10}$~GeV. Although this could allow for
interesting experimental tests of low Planck scale theories in a far future,
this remains extremely far above the currently highest energies measured
($\sim 100$~GeV for antiprotons).\\

Another potential concern about the $\mu$BHs production compatibility with data
is related to the flux of emitted gamma-rays. They result both from the direct 
emission and from the decay of neutral
pions. As the number of partonic degrees of freedom dominates over the
electromagnetic ones in the standard model, this latter component is clearly
dominant. It is much higher than the antiproton emission (as pions are much more
numerous in the fragmentation process) but so is the astrophysical background.
The resulting gamma-ray spectrum has been estimated using the PYTHIA generator 
to compute the pion distribution and there decay. It has then been Lorentz transformed
and integrated over the matter distribution of the Galaxy. It has also been
checked, as in \citep{uhecr}, that the extragalactic contribution is much lower
than the galactic one. The gamma-ray spectrum due to $\mu$BHs
remains lower than the background up to energies
of $10^{15}$~GeV (this latter being extrapolated with a power law above the
currently measured energy range). Such
energies, far above the GZK cutoff, are meaningless, especially when considering
that the primary cosmic-ray component necessary to produce the related $\mu$BHs should 
have much higher energies.\\

The conclusion of this study is that whatever the considered emitted cosmic-rays,
$\mu$BHs that would be currently produced in the Universe if the Planck scale is
around a TeV are fully compatible with astrophysical experimental data in spite of
a production rate around $10^{10}$~GHz per galaxy.

\section{Thermal production of black holes in the early Universe and relics}

It could also be thought that the possible thermal production of $\mu$BHs in the
early Universe should conflict with observations. If the temperature of the
Universe reached
values above the D-dimensional Planck scale $M_D$, $\mu$BHs are indeed expected 
to have formed through the
scattering of the thermal radiation. The production and subsequent evaporation
of such small black holes is irrelevant as long as it takes place before the
nucleosynthesis~: it just contributes to the thermal bath. On the other hand, if
black holes mostly decay after the nucleosynthesis, the entropy released could
modify the relative abundances of light elements and contradict the observational 
data (as it is well known for the gravitino problem in inflationary cosmology).
The mass of a D-dimensional $\mu$BH with lifetime $\tau$ is given by~:
$$M_{BH}=M_D\left( M_D \tau \right)^{\frac{D-3}{D-1}}.$$
Depending on the number of extra dimensions, the energy required to produce black holes
that will survive during $\sim 100$~s ({\it i.e.} until the nucleosynthesis)
is of the order of $10^{20}-10^{23}$~GeV, which is
above the usual quadri-dimensional Planck scale and clearly meaningless. It can
therefore be safely considered that $\mu$BHs thermally created in the early Universe 
are not expected to induce any substantial changes in the sensitive primordial
$^6Li$, $^3He$, $^4He$ and $D$ abundances.\\

A last point to address is the possible formation of stable relics at the end of the
evaporation process. Based on several different arguments ({\it e.g.} Gauss-Bonnet
solutions in string gravity \citep{relics1} or a renormalization group modification of 
the Schwarzschild metric \citep{relics2}), it has recently been pointed out that, in
agreement with the "cosmic censorship" hypothesis, black hole relics are expected to
survive the Hawking evaporation, even in models with extra-dimensions. In the Planck 
region, the semi-classical approach cannot be safely used anymore and quantum
gravitational effects must be taken into account. Although no reliable theory is yet
established, most attempts trying to deal with the endpoint of the evaporation
process lead to the conclusion that stable relics should be formed. Depending on the
details of the underlying model, their mass is expected to lie around the Planck mass. The possible
contribution of those relics to the cold dark matter budget has been extensively
studied in the framework of 4-dimensional black holes - see, {\it e.g.}, \citep{ann1,ann2} -
but it should be investigated whether the D-dimensional black holes produced by 
cosmic-ray interactions on the ISM could lead to a substantial amount of remnants.
The total mass induced by this process in our Galaxy can be computed by integrating
the formula given at the beginning of section 2~:
$$M_{relics}^{tot}=\int_{M_D}^{\infty}\int_{t_{form}}^{t_0}\int_0^{R_{gal}}\int_{0}^{y_{gal}}
2\pi RM_{rel}\frac{dN}{dM_{BH}dtdV}dM_{BH}dtdRdy$$
where $t_0$ is the age of the Universe, $t_{form}$ is the galaxy formation epoch,
$R_{gal}$ and $y_{gal}$ are the Milky-Way radius and height and $M_{rel}\sim M_D$ is
the relics expected mass. This leads to a tiny value $M_{relics}^{tot} \approx
10^{12}$~g which means that relics dark-matter is negligible and therefore remains compatible
with observations.

\section*{Conclusion and prospects}

This study shows that the concern related with $\mu$BHs created by astrophysical processes
(either in the Galaxy or in the primordial Universe) which could conflict with experimental
data (either through the emitted
cosmic-rays or through the possible relics) is irrelevant. From the point of view 
of black hole formation, the earth atmosphere and
colliders are probably the only places where a TeV Planck scale should have
observational consequences. Theoretical motivations for extra-dimensional models
are therefore not currently contradicted by astroparticle physics phenomenology.
On the other hand, some space is still open for new phenomena
due to $\mu$BH evaporation in the Galaxy~: as their production cross-section has no
suppression factor and as their coupling with any quantum during the evaporation process
is purely gravitational, one can expect an abundant production of rare
particles. Furthermore, because of the large center-of-mass velocity with respect to the
galactic frame, those particles
are expected to lie in the high energy range with a hard spectrum. Although this
would make a significant difference with the usual background thermal distributions, a clear
detection seems quite doubtful and would require a specific investigation.

\bigskip

{\it Aknowledgement}~: The authors would like to thank Gilles Henri whose
remarks where at the origin of this work.

\clearpage

\clearpage

\end{document}